\documentclass{sig-alternate}
\usepackage{url}
\usepackage{amssymb}
\usepackage{enumitem}
\usepackage{pifont}

\newcommand{\argmax}[1]{\underset{#1}{\operatorname{arg}\,\operatorname{max}}\;}

\usepackage{multirow}
\usepackage{array}
\usepackage{placeins}
\usepackage{balance}
\usepackage{rotating}
\usepackage{booktabs}
\usepackage{rotating}
\usepackage{hhline}
\usepackage{hyphsubst}
\usepackage{amsmath}
\usepackage{graphicx}
\usepackage{caption}
\usepackage{subcaption}
\usepackage{times}
\usepackage{ctable}

\def\hlinewd#1{%
\noalign{\ifnum0=`}\fi\hrule \@height #1 %
\futurelet\reserved@a\@xhline}

\begin{document}
\title{Long Time No See:\\
The Probability of Reusing Tags as a Function of Frequency and Recency}

\numberofauthors{4}
\author{
Dominik Kowald\\
    \affaddr{Know-Center}\\
       \affaddr{Graz, Austria}\\
       \email{dkowald@know-center.at}
\alignauthor
Paul Seitlinger\\
       \affaddr{Graz University of Technology}\\
       \affaddr{Graz, Austria}\\
       \email{paul.seitlinger@tugraz.at}
       \and
       \alignauthor
Christoph Trattner\\
       \affaddr{Know-Center}\\
       \affaddr{Graz, Austria}\\
       \email{ctrattner@know-center.at}
\alignauthor
Tobias Ley\\
       \affaddr{Tallinn University}\\
       \affaddr{Tallinn, Estonia}\\
       \email{tley@tlu.ee}
}

\maketitle
\begin{abstract}
In this paper, we introduce a tag recommendation algorithm that mimics the way humans draw on items in their long-term memory. This approach uses the frequency and recency of previous tag assignments to estimate the probability of reusing a particular tag. Using three real-world folksonomies gathered from bookmarks in BibSonomy, CiteULike and Flickr, we show how adding a time-dependent component outperforms conventional "most popular tags" approaches and another existing and very effective but less theory-driven, time-dependent recommendation mechanism. By combining our approach with a simple resource-specific frequency analysis, our algorithm outperforms other well-established algorithms, such as FolkRank, Pairwise Interaction Tensor Factorization and Collaborative Filtering. We conclude that our approach provides an accurate and computationally efficient model of a user's temporal tagging behavior. We show how effective principles for information retrieval can be designed and implemented if human memory processes are taken into account.
\end{abstract}



\section{Introduction} \label{sec:introduction}
One of the goals of Web Science as a new discipline is to understand the dynamics of human behavior and social interactions that shape the Web as a vast information network of content and people. As the Web evolves into a platform through which people interact with each other, communicate and express themselves, models of human behavior can shed light on why the Web forms as it does, and at the same time can contribute to improve underlying mechanisms of how the Web works. 

In this paper, we suggest a tag recommendation mechanism that mimics how people access their memory to name things they encountered in the past. In everyday communication people are very effective and quick in retrieving relevant knowledge from the enormous amount of information units stored in their individual long-term memory (LTM). One example is tagging resources on the Web, a rudimentary variant of communication \cite{Halpin2007, steels06IEEE}. Here, people name objects, such as images or music files, by means of social tags to create retrieval cues for personal and collective information organization \cite{Marlow2006}. 
The question how human memory ensures a fast and automatic information retrieval from its huge LTM has been insightfully examined in memory psychology (e.g., \cite{Anderson2004}). 
Briefly speaking, human memory is tuned to the statistical structure of an individual's environment and keeps available those memory traces that have been used frequently and recently in the past \cite{anderson_reflections_1991}.

Social tagging provides an illustrative example of the strong interplay between external, environmental and internal memory structures (e.g., \cite{Held201234}). For instance, the development of generative models of social tagging has revealed that the probability of a tag being applied can be modeled through the preferential attachment principle (e.g., \cite{Dellschaft2008}): the higher the frequency of a tag's past occurrence within the whole tagging environment, the more likely it is to be reused by an individual. Additionally, the same probability is also a function of the tag's recency, which is the time elapsed since the tag last occurred in the environment \cite{Cattuto30012007}. Summing up, the probability of applying a particular word reflects the individual's probability of being exposed to the word in her environment \cite{anderson_reflections_1991}.

The base-level learning (BLL) equation of the cognitive architecture ACT-R (e.g., \cite{Anderson2004}) combines the variables of frequency and recency of item exposure to estimate the base-level activation BLA$_i$ of the memory trace $i$ for the corresponding item. It is given by
  \begin{align}
		BLA_i = ln(\sum\limits_{j = 1}\limits^{n}{t_{j}^{-d}})
  \end{align}
, where $n$ represents the frequency of item occurrences in the past and $t_j$ symbolizes the recency, which is the time since the $j^{th}$ occurrence. The exponent $d$ accounts for the power-law of forgetting and models the phenomenon that each memory's activation, caused by the $j^{th}$ occurrence, decreases in time according to a power function. The exponent $d$ is typically set to $0.5$ \cite{Anderson2004}. 

Referring to the research briefly described above, we assume that a user's past usage of a tag predicts the probability that she will use the tag again in future. Hence, equation (1) should help to infer the probability of a tag being applied during a new tag assignment. In particular, if frequency and recency are both strong predictors of a tag's reuse probability, the base-level learning equation should help to extend simple "most popular tags" approaches that are only based on frequency analyses.

The work of \cite{zhang2012integrating} provides empirical evidence for this assumption. They showed that a recommender, which combines frequency and recency of tag use, reaches higher accuracy with respect to recall and precision than a recommender only taking into account the frequency of tag use. However, the equations they used to implement their approach were developed from scratch and not derived from existing research described above (see Sections \ref{sec:algos} and \ref{sec:dis_con}).

The research questions of this work are as follows: (i) Does the BLL equation provide a valid model of a user's tagging behavior in the past to predict future tag assignments? (ii) Can the BLL equation, that integrates frequency and recency of tag usage, be applied and extended to realize an effective and efficient tag recommendation mechanism?

The strategy we chose to address both research questions consisted of two steps. In a first step, we implemented the "pure" BLL equation in form of a tag recommender 
and compared its performance with a MostPopular$_u$ (MP$_u$) approach suggesting the most frequent tags in a user's tag assignments. This comparison should reveal the increment value that may result from additionally processing the recency of tag use. Moreover, we compared our BLL recommender with the approach introduced by \cite{zhang2012integrating} in order to reveal potential advantages of our theory-driven approach. 

In a second step, we extended the BLL equation to also take into account the effect of popular tags (i.e., semantic cues $C$) associated with a resource on the availability of memory traces and hence, tagging behavior. As a first approximation of $C$ we decided to simply weight the tags based on their frequency in the tag assignments of the resource (hereinafter called MostPopular$_r$ (MP$_r$)). We then compared the performance of this combination of BLL and MP$_r$ (BLL+C) with well-established approaches, such as Collaborative Filtering (CF), FolkRank (FR) and Pairwise Interaction Tensor Factorization (PITF), to examine our second research question. 

The remainder of this paper is organized as follows. We begin with discussing related work (Section \ref{sec:related_work}) and describing our approach in Section \ref{sec:approach}. Sections \ref{sec:experimental_setup} and \ref{sec:results} address our two research questions and summarize the settings and results of our extensive evaluation. Finally, in Section \ref{sec:dis_con}, we conclude the paper by discussing our findings in the light of the benefits of deriving tag recommender mechanisms from empirical, cognitive research. 

\section{Related Work} \label{sec:related_work}
Recent years have shown that tagging is an important feature of the Social Web
supporting the users with a simple mechanism to collaboratively organize and finding content \cite{Korner2010}.
Although tagging has been shown to significantly improve search \cite{Dellschaft2012} (and in particular tags provided by the individual), it is also known that users
are typically lazy in providing tags for instance for their bookmarked resources. It is therefore not surprising that recent research has taken up this challenge to support
the individual in her tag application process in the form of personalized tag recommenders. To date, the two following approaches have been established -- 
graph based and content-based tag recommender systems \cite{lipczak2012hybrid}. In our work we focus on graph-based approaches.


The probably most notable work in this context is the work of Hotho et al. \cite{hotho2006information} who introduce an algorithm called FolkRank (FR) that has  established itself as the most prominent benchmarking tag recommender approach over the past few years. Subsequent work in this context is the work of J\"{a}schke et al. \cite{jaschke2007tag} or Hamouda \& Wanas \cite{hamouda2011put} who show how the classic Collaborative Filtering (CF) approach could be adopted for the problem of predicting tags to the user in a personalized manner. More recent work in this context are studies of
Rendle et al. \cite{Rendle2010}, Wetzker et al. \cite{wetzker2010tag}, Krestel et al. \cite{krestel2010language} or Rawashdeh et al. \cite{rawashdeh2012folksonomy} who introduce a factorization model, a Latent Dirichlet Allocation (LDA) model or a Link-Prediction model, based on the Katz measure, to recommend tags to users. 

Although the latter mentioned approaches perform reasonable well, they are computational expensive compared to simple "most popular tags" approaches. Furthermore, they ignore recent observations made in social tagging systems, such as the variation of the individual tagging behavior over time \cite{yin2011temporal}. To that end, recent research has made first promising steps towards more accurate graph-based models that also account for the variable of time \cite{yin2011exploiting,zhang2012integrating}. The approaches have shown to outperform some of the current state-of-the-art tag recommender algorithms. 

Related to the latter strand of research, we present in this paper a novel graph-based tag recommender mechanism that uses the BLL equation which is based on the principles of a popular model of human cognition called ACT-R (e.g., \cite{Anderson2004}). We show that the approach is not only extremely simple but also reveal that the algorithm outperforms current state-of-the-art graph-based (e.g., \cite{wetzker2010tag,hotho2006information,jaschke2007tag}) and the leading time-based \cite{zhang2012integrating} tag recommender approaches.

\section{Approach} \label{sec:approach}
In Section \ref{sec:introduction} we formulated the assumption that both frequency and recency of tag use explain a large amount of variance in a tag's probability being applied and that this probability can be modeled through the BLL equation introduced by Anderson et al. \cite{Anderson2004}. In the following we describe how we have implemented the BLL equation to calculate the base-level activation (BLA) of a given tag $t$ in a user's set of tag assignments, $Y_{t, u}$. First, we determined a reference timestamp $timestamp_{ref}$ (in seconds) that is the timestamp of the most recent bookmark of user $u$. In our dataset samples, $timestamp_{ref}$ corresponded to the timestamp of $u$'s bookmark in the test set (see Section \ref{sec:dataset}). 
If $i$ = 1 ... $n$ index all tag assignments in $Y_{t, u}$, the recency of a particular tag assignment is given by $timestamp_{ref} - timestamp_i$. Finally, the BLA of tag $t$ for a user $u$ is given by the BLL equation:
	\begin{align}
		BLA(t, u) = ln(\sum\limits_{i = 1}\limits^n{(timestamp_{ref} - timestamp_{i})^{-d})}
  \end{align}
, where $d$ is set to 0.5 based on \cite{Anderson2004}. In order to map the values onto a range of 0 - 1 we applied a normalization method as proposed in related work \cite{mcauleyhidden}: 
	\begin{align}
		\|BLA(t,u)\| = \frac{exp(BLA(t, u))}{\sum\limits_{t' = 1}\limits^{m}{exp(BLA(t', u))}}
  \end{align}
, where \textit{m} equals $|Y_{u}|$.  

When incorporating BLL into a recommender, we aim at predicting the probability of a word being applied in the present tag assignment. To this end, we also have to take into account semantic cues $C$ in a user's current environment (e.g., the resource to be tagged) to fine-tune the "prior" probability estimated by means of the BLL equation (e.g., \cite{Anderson2004}). In case of tagging a resource, $C$ partially consists of content words in the title and in the page text or of prominent tags associated with the resource (e.g., \cite{Lorince2013, lipczak2012hybrid}). Since we focus in this work on graph-based approaches and not all of our datasets contain title information nor page-text, we 
modeled the influence of $C$ by simply taking into account the most popular tags of the resource (MP$_r$ , i.e., $\argmax{t \in T}(|Y_{t, r}|))$ \cite{hotho2006information}. Thus, we applied MP$_r$ to adjust the BLA of a given tag according to potential semantic cues available in the user's environment. 
Taken together, the list of recommended tags for a given user $u$ and resource $r$ is calculated by
  \begin{align}
		\widetilde{T}(u, r) = \argmax{t \in T}(\underbrace{\beta \underbrace{\|BLA(t, u)\|}_{BLL} + (1 - \beta) \||Y_{t, r}|\|}_{BLL+C})
  \end{align}
, where $\beta$ is used to inversely weight the two components, i.e. the BLA and the semantic cues $C$. The results presented in Section \ref{sec:results} were calculated with $\beta$ = 0.5. However, we focused on the performance of BLL+C in the experiments, i.e. on an approach estimating a tag's probability being applied by means of user and resource information.
Taken together, this is in line with the ACT-R declarative module that also considers retrieval probability as a function of base-level activation and environmental features.

\begin{table}[t!]
\small
  \setlength{\tabcolsep}{2.6pt}
  \centering
    \begin{tabular}{lllllll}
    \specialrule{.2em}{.1em}{.1em}
      Dataset			& Core	& $|B|$			& $|U|$	& $|R|$	& $|T|$	& $|TAS|$	 \\ \hline 
      BibSonomy	  & -			& 400,983 & 5,488  	& 346,444		& 103,503	& 1,479,970						\\
									& 3		& 41,764 	& 788  		& 8,711			& 5,757		& 161,509							\\\hline

					
			CiteULike		&	-		& 3,879,371 & 83,225  & 2,955,132		& 800,052	& 16,703,839							\\
									& 3		& 735,292 & 17,983  & 149,220		& 67,072	& 2,242,849							\\\hline
															
			Flickr			& -		& 864,679 	& 9,590  	& 864,679		& 127,599		& 3,552,540							\\ 
									& 3		& 860,135 	& 8,332  	& 860,135			& 58,831		& 3,465,346							\\
									
		\specialrule{.2em}{.1em}{.1em}								
    \end{tabular}
    \caption{Properties of the datasets, where $|B|$ is the number of bookmarks, $|U|$ the number of users, $|R|$ the number of resources, $|T|$ the number of tags and $|TAS|$ the number of tag assignments.}
  \label{tab:dataset_stats}
\end{table}

\section{Experimental Setup} \label{sec:experimental_setup}
In this section we describe in detail the datasets, the evaluation method, the metrics and the algorithms used for our experiments.

\subsection{Datasets} \label{sec:dataset}
For the purpose of our study and for reasons of reproducibility we focused with our investigations on three well-known and freely-available folksonomy datasets. To test our approach on both types of advocates -- namely known as broad and narrow folksonomies \cite{Helic2012} (in broad folksonomy many users are allowed to annotate a particular resource while in a narrow folksonomy only the user who has uploaded the resource is permitted to apply tags), freely available datasets from the social bookmark and publication sharing system BibSonomy\footnote{\url{http://www.kde.cs.uni-kassel.de/bibsonomy/dumps}}, the reference management system CiteULike\footnote{\url{http://www.citeulike.org/faq/data.adp}} (broad folksonomies) and the image and video sharing platform Flickr\footnote{\url{http://www.tagora-project.eu/}} (narrow folksonomy) were utilized. Since automatically generated tags have an impact on the performance of the tag recommender systems, we excluded all of those tags from the datasets, e.g., for BibSonomy and CiteULike we excluded for instance the \textit{no-tag}, \textit{bibtex-import}-tag etc. Furthermore, we decapitalized all tags as suggested by related work in the field (e.g., \cite{Rendle2010}). In the case of Flickr we randomly selected 3\% of the user profiles for reasons of computational effort (see also \cite{gemmell2009improving}). The overall dataset statistics can be found in Table \ref{tab:dataset_stats}. As depicted, we applied both: a $p$-core pruning approach \cite{batagelj2002generalized} to capture the issue of data sparseness, as well as no $p$-core pruning to capture the issue of cold-start users or items \cite{Doerfel2013}.

\begin{table}[t!]
  \setlength{\tabcolsep}{7pt}
  \centering
  \small
    \begin{tabular}{llllll}
    \specialrule{.2em}{.1em}{.1em}
      Dataset		& Core	& Measure 	& MP$_u$ 				& GIRP 			  & BLL  					\\ \hline 
      					                                                
			BibSonomy	& -	& $F_1$@5   & .152  				& .157  			&	\textbf{.162} 						\\
								&		&	MRR			  & .114  				& .119  			&	\textbf{.125} 					\\
								&		&	MAP			  & .148  				& .155  			&	\textbf{.162} 				\\\cline{2-6}
								& 3	& $F_1$@5 	& .215  		& .221 		&	\textbf{.228} 				\\
								&		&	MRR				& .202  		& .210  	&	\textbf{.230} 				\\
								&		&	MAP				& .238  		& .247  	&	\textbf{.272}         			\\\hline                                                                
												                                        
%
			CiteULike	& -	& $F_1$@5	 	& .185 					& .194  			&	\textbf{.201} 										\\
								&		&	MRR			  & .165  				& .182  			&	\textbf{.193} 									\\
								&		&	MAP			  & .194  				& .213  			&	\textbf{.227} 				\\\cline{2-6}
								& 3	& $F_1$@5	& .272 	& .291  &	\textbf{.300}										\\
								&		&	MRR			& .268  & .294  &	\textbf{.319	}								\\
								&		&	MAP			& .305  & .337  &	\textbf{.366	}		\\\hline													                                      
													                                      
			Flickr		& -	& $F_1$@5	 	& .435  				& .509  			&	\textbf{.523} 										\\
								&		&	MRR			  & .360 					& .445  			&	\textbf{.466} 									\\
								&		&	MAP			  & .468  				& .590  			&	\textbf{.619} 			\\\cline{2-6}							
								& 3	& $F_1$@5	& .488  	& .577 &	\textbf{.592}										\\
								&		&	MRR			& .407 		& .511 &	\textbf{.533}									\\
								&		&	MAP			& .527  	& .676 &	\textbf{.707}			\\
		\specialrule{.2em}{.1em}{.1em}		
    \end{tabular}
    \caption{$F_1$@5, MRR and MAP values for BibSonomy, CiteULike and Flickr (no core and core 3) showing that our BLL equation provides a valid model of a user's tagging behavior to predict tags (first research question).}
  \label{tab:mrr_map_1_user}
\end{table}

\begin{table*}[t!]
  \setlength{\tabcolsep}{7pt}
  \centering
  \small
    \begin{tabular}{lllllllllllll}
    \specialrule{.2em}{.1em}{.1em}
      Dataset		& Core	& Measure & MP				& MP$_r$	& MP$_{u,r}$	 	& CF 		& APR		& FR 			& FM			& PITF		& GIRPTM   & BLL+C 									\\ \hline 
      					                                                                                                        
			BibSonomy	& -			& $F_1$@5 & .013  		& .074		& .192					& .166	& .175	& .171		&	.122		&	.139		& .197		 & \textbf{.201}						\\
								&				&	MRR			& .008  		& .054		& .148					& .133	& .149	& .148		&	.097		&	.120		& .152  	 & \textbf{.158}					\\
								&				&	MAP			& .009  		& .070		& .194					& .173	& .193	& .194		&	.120		&	.150		&	.200  	 & \textbf{.207}				\\\cline{2-13}
								& 3			& $F_1$@5 & .047  		& .313		& .335					& .325	& .260	& .337		&	.345		&	.356		& .350		 & \textbf{.353}									\\
								&				&	MRR			& .035  		& .283		& .327					& .289	& .279	& .333		&	.329		&	.341		& .334  	 & \textbf{.349}								\\
								&				&	MAP			& .038  		& .345		& .403					& .356	& .329	& .414		&	.408		&	.421	  &	.416  	 & \textbf{.435}			\\\hline                                                                                                                        
																																																														
%
			CiteULike	& -			& $F_1$@5	& .002 			& .131		& .253					& .218	& .195 	& .194		&	.111		&	.122		& .263		 & \textbf{.270}										\\
								&				&	MRR			& .001  		& .104		& .229					& .201	& .233 	& .233		&	.110		&	.141		& .246		 & \textbf{.258}									\\
								&				&	MAP			& .001  		& .134		& .280					& .247	& .284 	& .284		&	.125		&	.158		& .301		 & \textbf{.315}				\\\cline{2-13}
								& 3			& $F_1$@5	& .013 			& .270		& .316					& .332	& .313 	& .318		&	.254		&	.258		& .336		 & \textbf{.346}										\\
								&				&	MRR			& .012  		& .243		& .353					& .295	& .361 	& .366		&	.282		&	.290		& .380		 & \textbf{.409}									\\
								&				&	MAP			& .012  		& .294		& .420					& .363	& .429 	& .436		&	.326		&	.334		& .455		 & \textbf{.489}			\\\hline																                                                                                        
																																																														
			Flickr		& -			& $F_1$@5	& .023 			& -				& .435					& .417	& .328 	& .334		&	.297		&	.316		& .509		 & \textbf{.523}										\\
								&				&	MRR			& .023  		& -				& .360					& .436	& .352	& .355		&	.300		&	.333		& .445		 & \textbf{.466}									\\
								&				&	MAP			& .023  		& -				& .468					& .581	& .453 	& .459		&	.384		&	.426		& .590		 & \textbf{.619}					\\\cline{2-13}
								& 3			& $F_1$@5	& .026 	  	& -				& .488					& .493	& .368 	& .378		&	.361		&	.369		& .577		 & \textbf{.592}										\\
								&				&	MRR			& .026   		& -				& .407					& .498	& .398	& .404		&	.375		&	.390		& .511		 & \textbf{.533}									\\
								&				&	MAP			& .026    	& -				& .527					& .663	& .513 	& .523		&	.481		&	.502		& .676		 & \textbf{.707}									\\										
		\specialrule{.2em}{.1em}{.1em}		
    \end{tabular}
    \caption{$F_1$@5, MRR and MAP values for BibSonomy, CiteULike and Flickr (no core and core 3) showing that BLL+C outperforms state-of-the-art baseline algorithms (second research question).}
  \label{tab:mrr_map_1}
\end{table*}

\subsection{Evaluation Methodology} \label{sec:evaluation_metrics}
To evaluate our tag recommender approach we used a leave-one-out hold-out method as proposed by popular and related work in this area (e.g., \cite{jaschke2007tag}). Hence, we created two sets, one for training and one for testing. To split up each dataset in those two sets we eliminated for each user her latest bookmark (in time) from the original dataset and added it to the test set. The remaining original dataset was then used for training, and the newly created one for testing. This procedure simulates well a real-world environment and is a recommended offline-evaluation procedure for time-based recommender systems \cite{campos2013time}.
To finally quantify the performance of our approach, a set of well-known information retrieval performance standard metrics were used . In particular, we report Recall (R@$k$), Precision (P@$k$), F1-Score ($F_1$@$k$), Mean Reciprocal Rank (MRR) and Mean Average Precision (MAP), where $k$ is between 1 and 10 and MRR and MAP are calculated for 10 recommended tags ($k$ = 10) \cite{jaschke2007tag, lipczak2012hybrid}.

\subsection{Baseline Algorithms}\label{sec:algos}
We compared the results of our approach to several "baseline" tag recommender algorithms. The algorithms were selected in respect to their popularity in the community, performance and novelty \cite{recbook}:

\textbf{MostPopular (MP):}
This approach recommends for any user and any resource the same set of tags that is weighted by the frequency in all tag assignments \cite{jaschke2008tag}.

\textbf{MostPopular$_u$ (MP$_u$):}
The \textit{most popular tags by user} approach suggests the most frequent tags in the tag assignments of the user \cite{jaschke2008tag}.
%

\textbf{MostPopular$_r$ (MP$_r$):}
The \textit{most popular tags by resource} algorithm weights the tags based on their frequency in the tag assignments of the resource \cite{jaschke2008tag}.
	
\textbf{MostPopular$_{u, r}$ (MP$_{u,r}$):}
This algorithm is a mixture of the most popular tags by user (MP$_u$) and most popular tags by resource (MP$_r$) approaches \cite{jaschke2007tag}.

\textbf{Collaborative Filtering (CF):}
Marinho et al. \cite{marinho2008collaborative} described how the classic Collaborative Filtering (CF) approach \cite{schafer2007collaborative} can be adapted for tag recommendations, 
where the neighborhood of an user is formed based on the tag assignments in the user profile.
 The only variable parameter here is the number of users in the neighborhood which has been set to 20 based on the work of Gemmell et al. \cite{gemmell2009improving}.

\textbf{Adapted PageRank (APR):}
Hotho et al. \cite{hotho2006information} adapted the well-known PageRank algorithm in order to rank the nodes within the graph structure of a folksonomy. This is based on the idea that a resource is important if it is tagged with important tags by important users.

\textbf{FolkRank (FR):}
The FolkRank algorithm is an extension of the Adapted PageRank approach that 
gives a higher importance to the preference vector using a differential approach \cite{jaschke2007tag}. Our APR and FR implementations are based on the code and the settings of the open-source Java tag recommender framework provided by the University of Kassel\footnote{\url{http://www.kde.cs.uni-kassel.de/code}}. 

\textbf{Factorization Machines (FM):}
Rendle \cite{rendle2010factorization} introduced Factorization Machines which combine the advantages of Support Vector Machines (SVM) with factorizition machines to build a general prediction model that is also capable of tag recommendations.

\textbf{Pairwise Interaction Tensor Factorization (PITF):}
This approach proposed by Rendle and Schmidt-Thieme \cite{Rendle2010} is an extension of factorization models based on the Tucker Decomposition (TD) model that explicitly models the pairwise interactions between users, resources and tags. 
The FM and PITF results presented in this paper were calculated using the open-source C++ tag recommender framework provided by the University of Konstanz\footnote{\url{http://www.informatik.uni-konstanz.de/rendle/software/tag-recommender/}} with 256 factors as suggested in \cite{Rendle2010}.

\textbf{Temporal Tag Usage Patterns (GIRP):}
This time-dependent tag-recommender algorithm was presented by Zhang et al. \cite{zhang2012integrating} and is based on the frequency and the temporal usage of a user's tag assignments. In contrast to BLL it models the temporal tag usage with an exponential distribution and not a power-law distribution. Furthermore, it is only based on the first- and last-time usage of the tag and not by all of its usages.

\textbf{GIRP with Tag Relevance to Resource (GIRPTM):}
This is an extension of the GIRP algorithm that also takes the resource component into account as it is also done in BLL+C \cite{zhang2012integrating}.

\begin{figure*}[t!]
        \centering
        \begin{subfigure}[b]{0.3\textwidth}
                \centering
                \includegraphics[width=\textwidth]{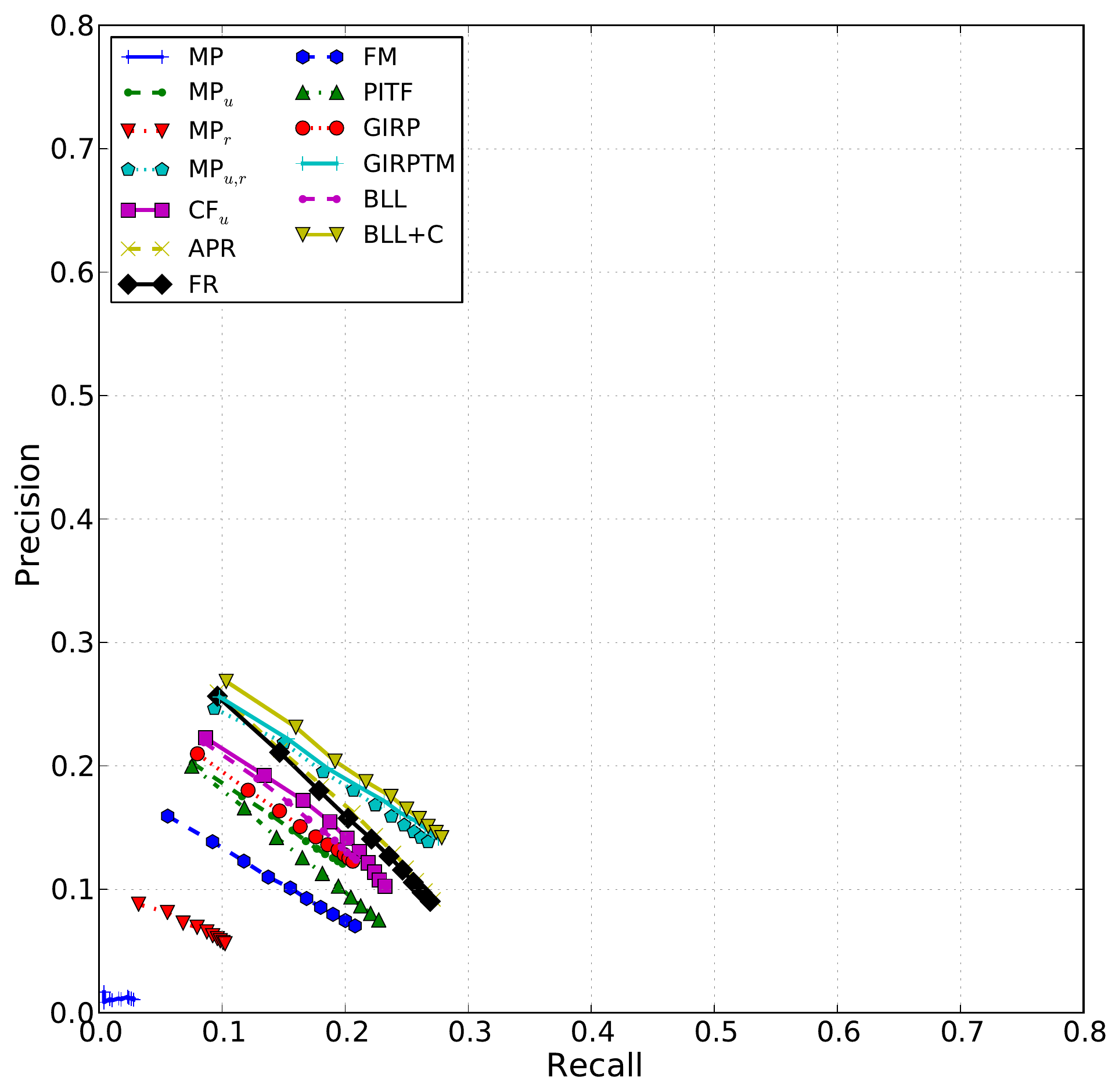}
                 \caption{BibSonomy (no core)}
                \label{fig:bib1_precrec}
        \end{subfigure}
          \begin{subfigure}[b]{0.3\textwidth}
                \centering
                \includegraphics[width=\textwidth]{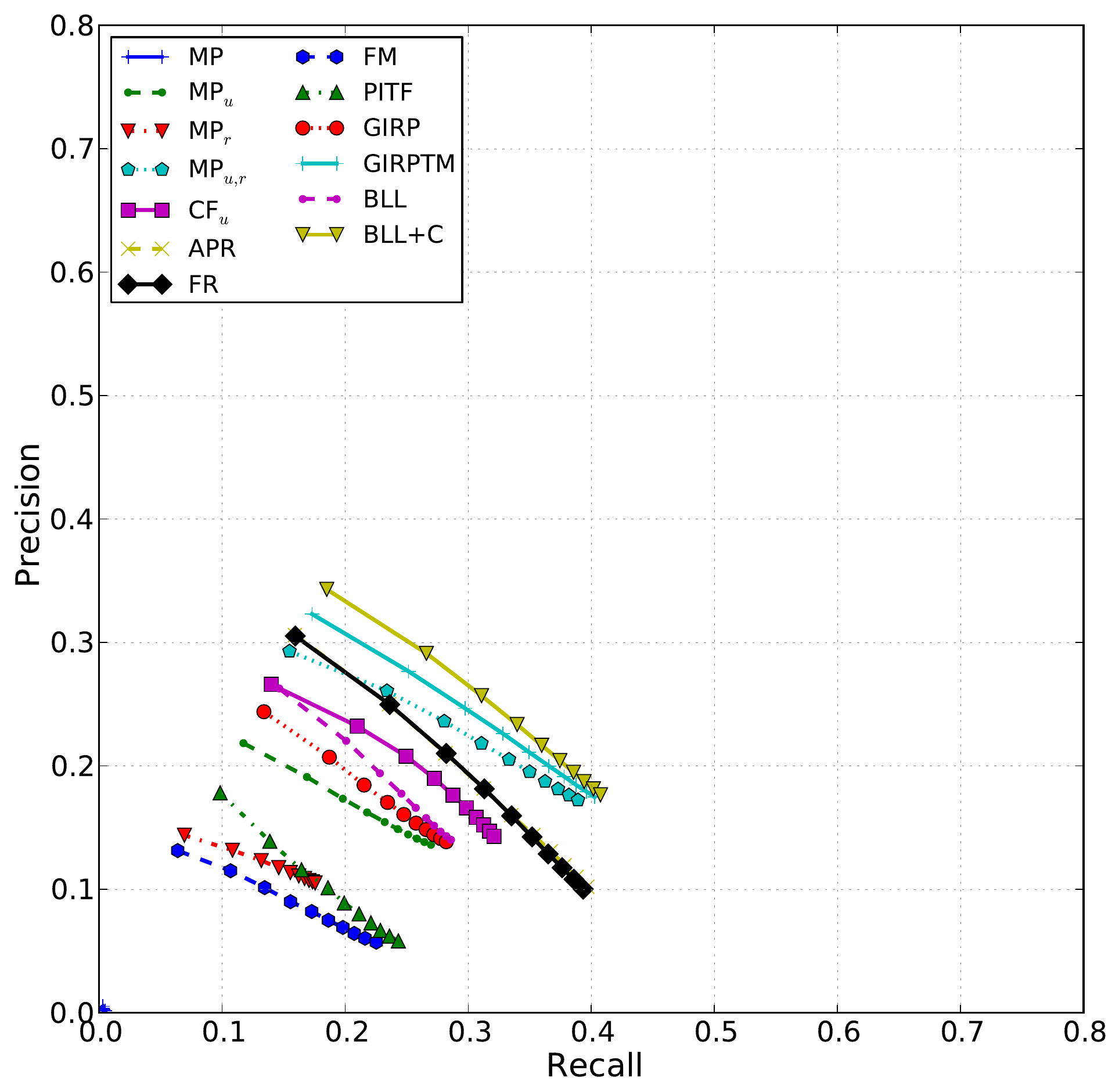}
                \caption{CiteULike (no core)}
                \label{fig:cul1_precrec}
        \end{subfigure}%
         \begin{subfigure}[b]{0.3\textwidth}
                \centering
                \includegraphics[width=\textwidth]{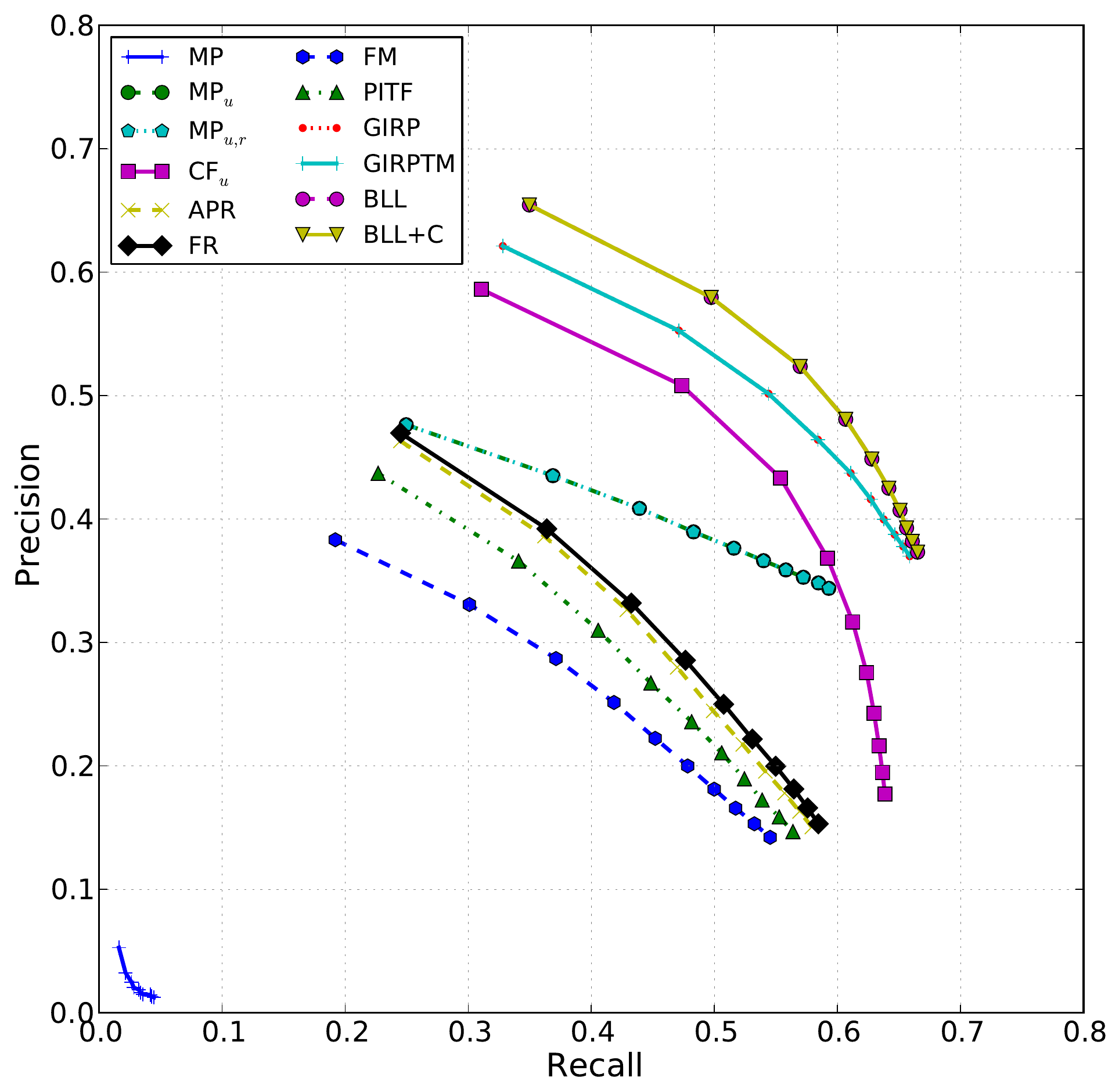}
                \caption{Flickr (no core)}
                \label{fig:flickr1_precrec}
        \end{subfigure}%
      
        \begin{subfigure}[b]{0.3\textwidth}
                \centering
                \includegraphics[width=\textwidth]{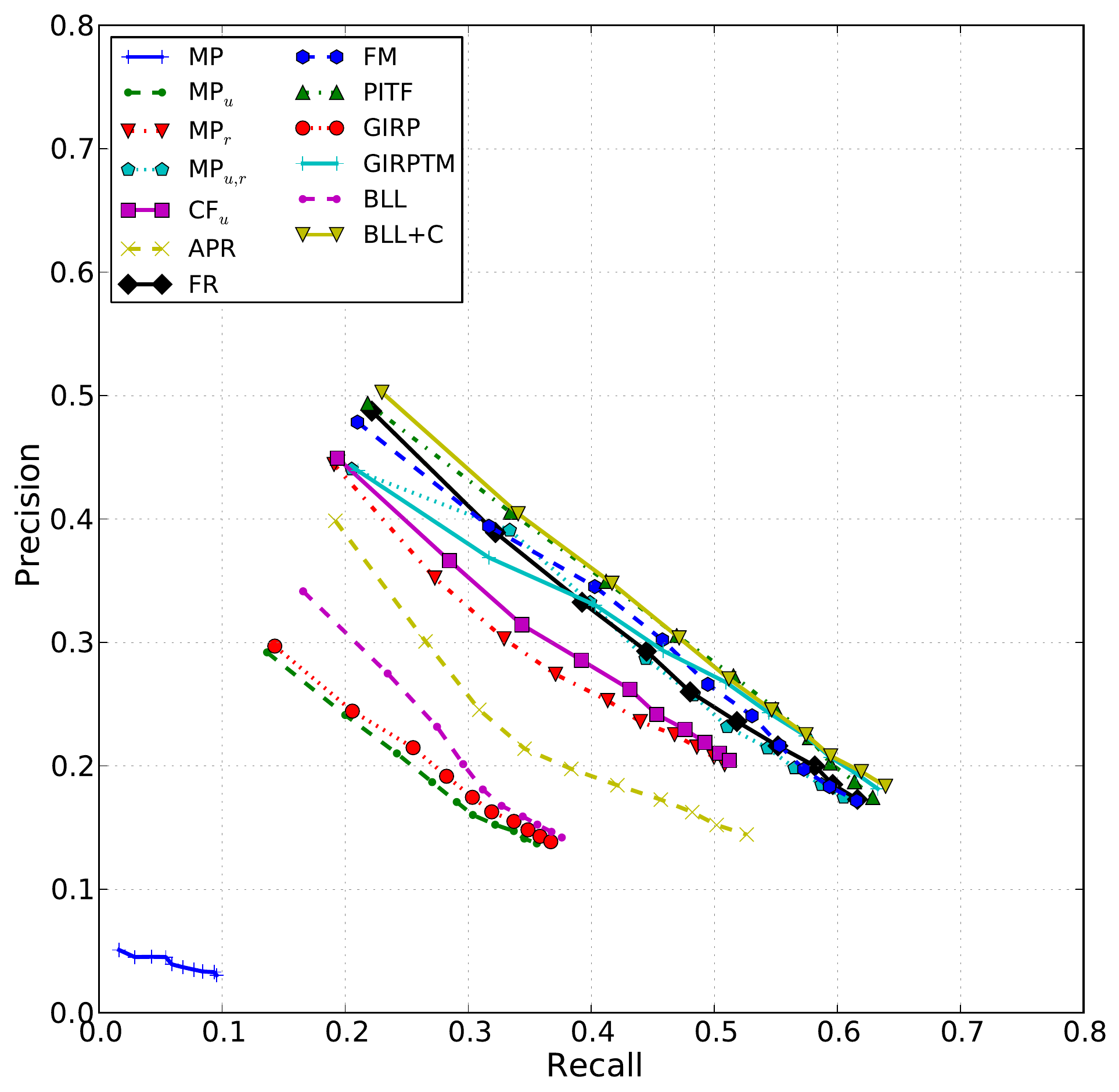}
                 \caption{BibSonomy (core 3)}
                \label{fig:bib3_precrec}
        \end{subfigure}
          \begin{subfigure}[b]{0.3\textwidth}
                \centering
                \includegraphics[width=\textwidth]{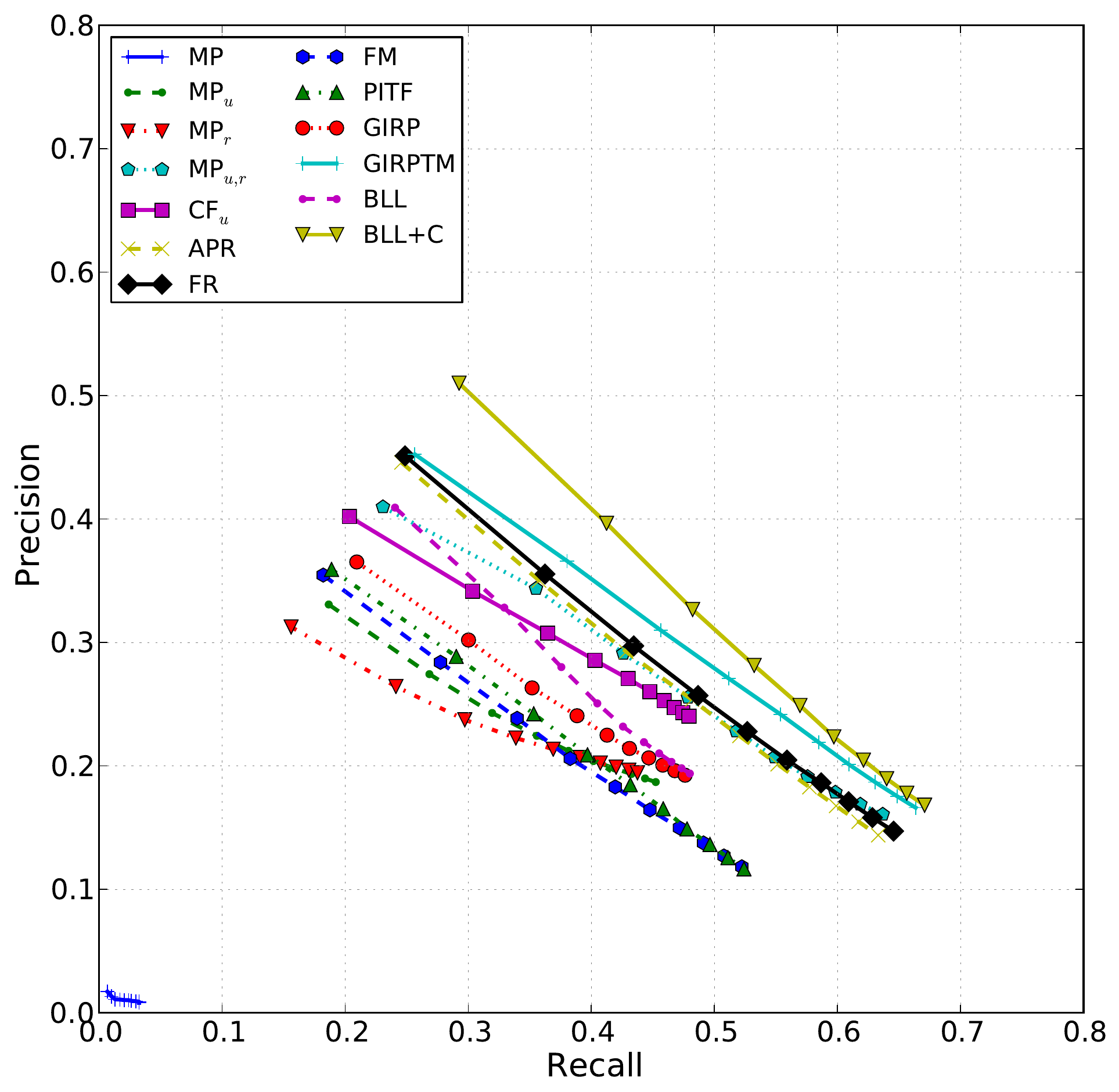}
                \caption{CiteULike (core 3)}
                \label{fig:cul3_precrec}
        \end{subfigure}%
         \begin{subfigure}[b]{0.3\textwidth}
                \centering
                \includegraphics[width=\textwidth]{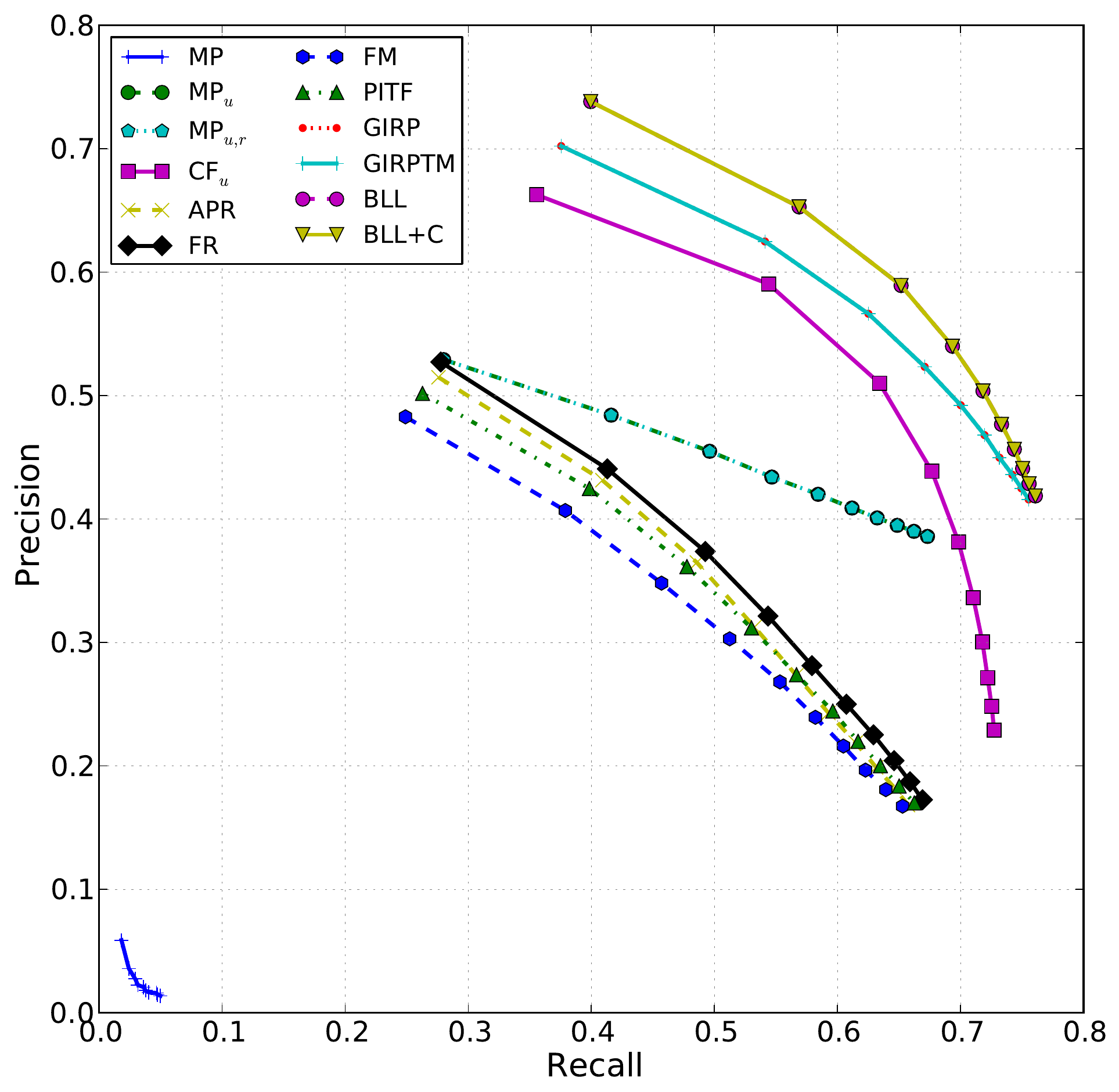}
                \caption{Flickr (core 3)}
                \label{fig:flickr3_precrec}
        \end{subfigure}%
        \caption{Recall/Precision plots for BibSonomy, CiteULike and Flickr (no core and core 3) showing the performance of the algorithms for 1 - 10 recommended tags ($k$).}
				\label{fig:prec_rec}
\end{figure*}

\section{Results} \label{sec:results}
In this section we present the results of our experiments in respect to the recommender quality in two steps as described in Section \ref{sec:introduction}. In the first step we compared BLL with MP$_u$ and GIRP in order to determine the impact of the recency component of the tag assignments. The results in Table~\ref{tab:mrr_map_1_user} clearly show that both time-dependent algorithms outperform the frequency-based MP$_u$ approach. Furthermore, BLL reaches higher levels of accuracy than the less theory-driven GIRP algorithm in both settings (without using a core and with core 3). Moreover, it becomes apparent that the impact of the recency component is significantly higher in the narrow folksonomy (Flickr) than it is in the broad folksonomies (BibSonomy, CiteULike).

In the second step we compared our BLL+C approach, which is a combination of BLL and MP$_r$, with a set of state-of-the-art baseline algorithms. When looking at the results in Table~\ref{tab:mrr_map_1}, the first thing that comes apparent is the fact that the two time-dependent algorithms (GIRPTM and BLL+C) reach the highest estimates ($F_1$@5, MRR and MAP) across all three datasets and both settings (with $p$-core pruning applied and without). Second, all recommender algorithms substantially outperform the baseline mechanism, i.e., the simplest "most popular tags" approach MP. Third, our BLL+C approach also outperforms GIRPTM, the currently leading graph-based time-depended tag recommendation algorithm, especially in terms of the ranking-dependent metrics, such as MRR and MAP. Same observations can be made when looking at the Recall / Precision curves in Figure~\ref{fig:prec_rec}.
%
%

Summing up, this pattern of results implies that the base-level learning equation can be used to implement a very effective recommender approach. By considering the recency in addition to frequency of tag use with the help of this equation as well as the current context, it exceeds the performance of well-established and effective recommenders, such as MP$_{u,r}$, CF, APR, FM and the other time-dependent approach GIRPTM. Most surprisingly, despite its simplicity, BLL+C appears to be even more successful than the sophisticated FR and PITF algorithms. 

The code we used for our experiments is open-source and can be found online\footnote{https://github.com/domkowald/tagrecommender}.

\section{Discussion and Conclusion} \label{sec:dis_con}
In this study we have followed a two-step strategy and started by comparing the performance of BLL with MP$_{u}$ to determine the effect of additionally considering the recency of each tag use as well as with GIRP to contrast our cognitive-psychological model with the less theory-driven approach introduced by Zhang et al. \cite{zhang2012integrating} in order to tackle our first research question. Our results clearly demonstrate that - independent of the evaluation metric and across all datasets - BLL reaches higher levels of accuracy than MP$_{u}$ and outperforms GIRP. Thus, processing the recency of tag use is effective to account for additional variance of users' tagging behavior and therefore, a reasonable extension of simple "most popular tags" approaches. Furthermore, the advantage over GIRP indicates that drawing on memory psychology guides the application of a reliable and valid model built upon long-standing, empirical research. The equations Zhang et al. \cite{zhang2012integrating} used to implement their approach were developed from scratch and not derived from existing research described above. As a consequence, \cite{zhang2012integrating} modeled recency of tag use by means of an exponential function that is clearly at odds with the power law of forgetting described in the introduction. Additionally, the model of \cite{zhang2012integrating} only considers the time elapsed since the first and last use of a tag and - in contrast to the BLL equation - does not take into account all other uses in between.

In a second step, we have combined BLL with MP$_r$. Where BLL gives the prior probability of tag reuse that is learned over time, MP$_r$ tunes this prior probability to the current context by exploiting the current semantic cues from the environment. This is in line with how ACT-R models the retrieval from long-term memory. Despite its simplicity, our results show that this combination (BLL+C) has potential to outperform well-established mechanisms, such as CF, FR and PITF. We assume this is the case because, in following some fundamental principles of human memory, BLL+C is better adapted to the statistical structure of the environment.

Moreover, a glance on the results shows an interaction between the dataset examined and the performance of BLL (and BLL+C). While the distance to other strongly performing mechanisms does not appear to be large in case of broad folksonomies (BibSonomy and CiteULike), this distance gets substantially larger in a narrow folksonomoy (Flickr). From this interaction we conclude that applying a model of human memory is primarily effective if tag assignments are mainly driven by individual habits unaffected by the behavior of other users, such as it is done in Flickr. 

In future work, we will continue examining memory processes involved in categorizing and tagging Web resources. For instance, in a recent study \cite{paul2013}, we have introduced a mechanism by which memory processes involved in tagging can be modeled on two levels of knowledge representation: on a semantic level (representing categories or LDA topics) and on a verbal level (representing tags). Next, we will aim at combining this integrative mechanism with the BLL equation to examine a potential interaction between the impact of recency (time-based forgetting) and the level of knowledge representation. Again, conclusions drawn from cognitive science should help to realize an effective and psychologically plausible tag recommendation mechanism.

\textbf{Acknowledgments:}
This work is supported by the Know-Center and the EU funded project Learning Layers (Grant Agreement 318209).

\balance
\bibliographystyle{abbrv}
\small
\bibliography{wsdm2014}
\balancecolumns
\end{document}